\documentclass[twocolumn,pra,showpacs,superscriptaddress]{revtex4-1}
\usepackage{amssymb}
\usepackage{amsmath}
\usepackage{graphicx}
\usepackage{epsfig}

\begin{document}

\title{Dynamical topological invariant for non-Hermitian Rice-Mele model}
\author{R. Wang}
\affiliation{School of Physics, Nankai University, Tianjin 300071, China}
\author{X. Z. Zhang}
\affiliation{College of Physics and Materials Science, Tianjin Normal University, Tianjin 300387, China}
\author{Z. Song}
\email{songtc@nankai.edu.cn}
\affiliation{School of Physics, Nankai University, Tianjin 300071, China}

\begin{abstract}
We study a non-Hermitian Rice-Mele model without breaking time-reversal
symmetry, with the non-Hermiticity arising from imbalanced hopping rates.
The Berry connection, Berry curvature and Chern number are introduced in the
context of biorthonormal inner product. It is shown that for a bulk system,
although the Berry connection can be complex numbers, the Chern number is
still quantized, as topological invariant. For an opened chain system, the
mid-gap edge modes are obtained exactly, obeying the bulk-edge
correspondence. Furthermore, we also introduce a local current in the
context of biorthonormal inner product to measure the pumping charge
generated by a cyclic adiabatic evolution. Analytical analysis and numerical
simulation of the time evolution of the mid-gap states show that the pumping
charge can be a dynamical topological invariant in correspondence with the
Chern number. It indicates that the geometric concepts for Hermitian
topological insulator can be extended to the non-Hermitian regime.
\end{abstract}

\maketitle



\section{Introduction}

Nowadays, non-Hermitian Hamiltonian is no longer a forbidden regime in
quantum mechanics since the discovery that a certain class of non-Hermitian
Hamiltonians could exhibit the entirely real spectra \cite{Bender,Bender1}.
In non-Hermitian quantum mechanics, the reality of the spectrum, unitary
evolution and probability conservation still hold if the Dirac inner product
is replaced by biorthonormal inner product. The origin of the reality of the
spectrum of a non-Hermitian Hamiltonian is the pseudo-Hermiticity of the
Hamiltonian operator \cite{Ali1}. Such kinds of Hamiltonians possess a
particular symmetry, i.e., it commutes with the combined operator $\mathcal{%
PT}$, but not necessarily with $\mathcal{P}$\ and $\mathcal{T}$\ separately.
Here $\mathcal{P}$\ is a unitary operator, such as parity, translation,
rotation operators etc., while $\mathcal{T}$\ is an anti-unitary operator,
such as time-reversal operator. The combined symmetry is said to be unbroken
if every eigenstate of the Hamiltonian is $\mathcal{PT}$-symmetric; then,
the entire spectrum is real, while is said to be spontaneously broken if
some eigenstates of the Hamiltonian are not the eigenstates of the combined
symmetry. However, even within the unbroken symmetric region,\ many
quantities, such as local particle probability, geometric phase of an
adiabatic evolution of an eigenstate, etc., exhibit anomalous behavior in
comparison with that in a Hermitian system. For instance, the biorthonormal
expectation value of a local particle number operator and the geometric
phase can be complex \cite{Bender2,F.G,D.P}. A natural question is that to
what extent a non-Hermitian system could inherit the property of a Hermitian
one by introducing biorthonormal inner product. In this paper, we
investigate a non-Hermitian RM model, the Hermitian counterpart of which is
a prototype system for topological matter.

We focus on the region without breaking time reversal symmetry, which
supports fully real spectrum. We investigate the topology of the degeneracy
point by introducing the concepts of Berry connection, Berry curvature and
Chern number in the context of biorthonormal inner product. In contrast to a
Hermitian system, the Berry connection can be complex numbers. However, the
Chern number is still quantized, as topological invariant to characterize
the feature of energy band. For an opened chain system, the mid-gap edge
modes are obtained exactly, that are identical to the ones in a Hermitian
system. The bulk-edge correspondence still hold in the corresponding
non-Hermitian SSH chain. Furthermore, we also introduce a local current in
the context of biorthonormal inner product to measure the pumping charge
generated by a cyclic adiabatic evolution, which is also defined by
biorthonormal probability. Analytical analysis and numerical simulation of
the time evolution of the mid-gap states show that the pumping charge can be
a dynamical topological invariant in correspondence with the Chern number.
Our results indicate that the geometric concepts for Hermitian topological
insulator can be extended to the non-Hermitian regime.

This paper is organized as follows. In Section \ref{Model Hamiltonian}, we
present the model Hamiltonian and the solutions. In Section \ref{Geometric
quantities}, we introduce the concepts of Berry connection, Berry curvature,
and Chern number for a non-Hermitian system. In Section \ref{Chern number},
we calculate the Chern number in the concrete system. In Section \ref%
{Edge-mode operators}, we study the edge modes for the open chain system.
Section \ref{Pumping charge} devotes to the dynamical signature of
topological feature of the degeneracy point. Finally, we give a summary and
discussion in Section \ref{Summary}.

\section{Model Hamiltonian}

\label{Model Hamiltonian}

We start our investigation by considering a non-Hermitian model with
imbalanced hopping%
\begin{equation}
H=\sum_{l=1}^{2N}[\kappa _{l,l+1}a_{l}^{\dagger }a_{l+1}+\kappa
_{l+1,l}a_{l+1}^{\dagger }a_{l}-V\left( -1\right) ^{l}a_{l}^{\dagger }a_{l}],
\end{equation}%
where $\kappa _{l,l+1}=\frac{1+\left( -1\right) ^{l}\delta }{2}\lambda
^{\left( -1\right) ^{l+1}}$, $\kappa _{l+1,l}=\frac{1+\left( -1\right)
^{l}\delta }{2}\lambda ^{\left( -1\right) ^{l}}$. The spinless fermions
satisfy the periodic boundary condition $a_{l}\equiv a_{l+2N}$. In this
paper, we focus on the non-Hermitian system, $\lambda \neq 1$ and $>0$. It
is a bipartite lattice, i.e., it has two sublattices $A$, $B$ such that each
site on lattice $A$ has its nearest neighbors on sublattice $B$, and vice
versa.\textbf{\ }The system is a variant of Hermitian SSH model by
introducing the imbalanced\ hopping and staggered real potentials. The
original Hermitian system at half-filling, proposed by Su, Schrieffer, and
Heeger to model polyacetylene \cite{Su,Schrieffer}, is the prototype of a
topologically nontrivial band insulator with a symmetry protected
topological phase \cite{Ryu,Wen}. In recent years, it has been attracted
much attention and extensive studies have been demonstrated \cite%
{Xiao,Hasan,Delplace,ChenS1,ChenS2,LS PRA}.

\begin{figure*}[tbp]
\includegraphics[ bb=37 371 416 681, width=0.3\textwidth, clip]{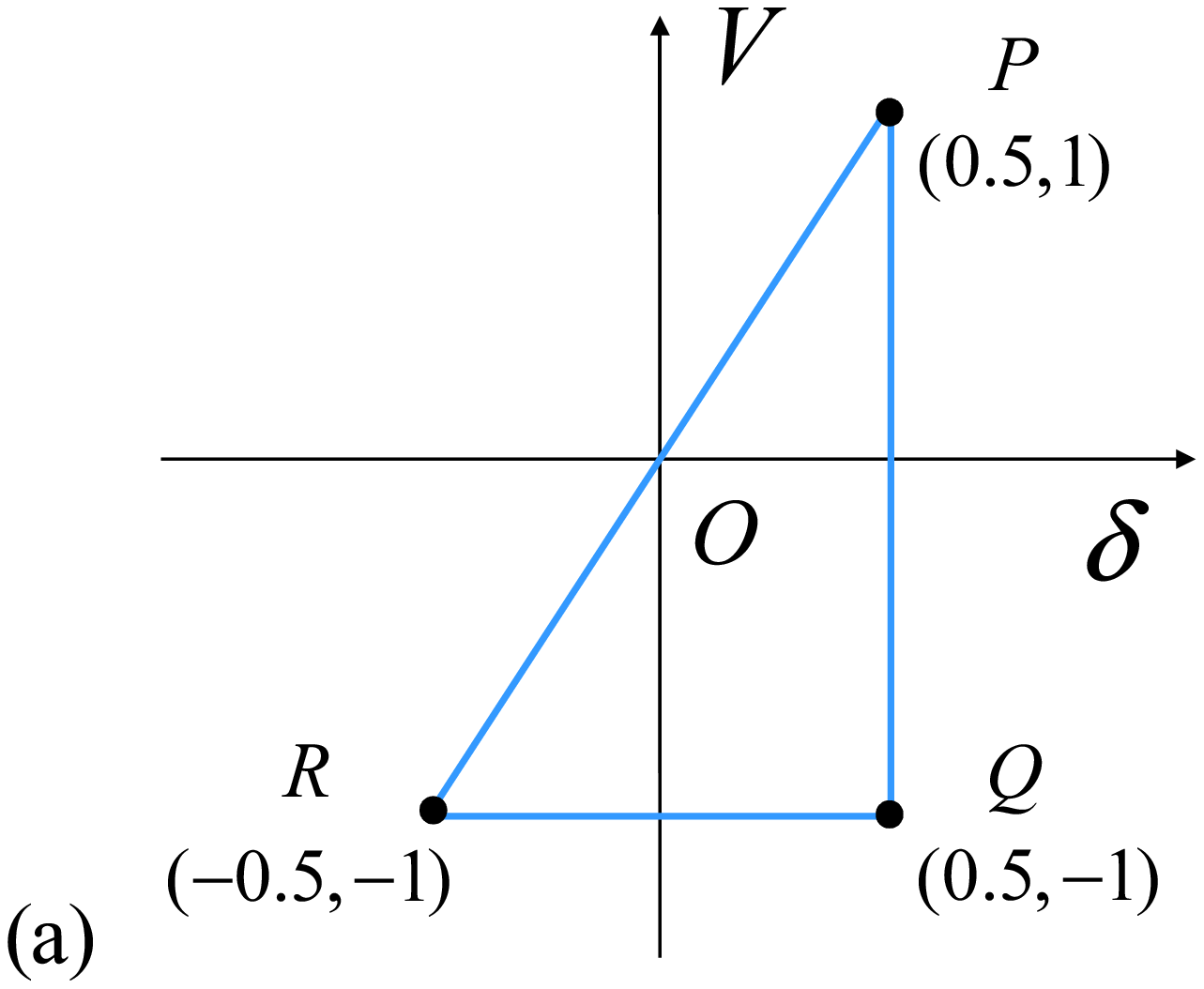} %
\includegraphics[ bb=89 236 514 553, width=0.3\textwidth, clip]{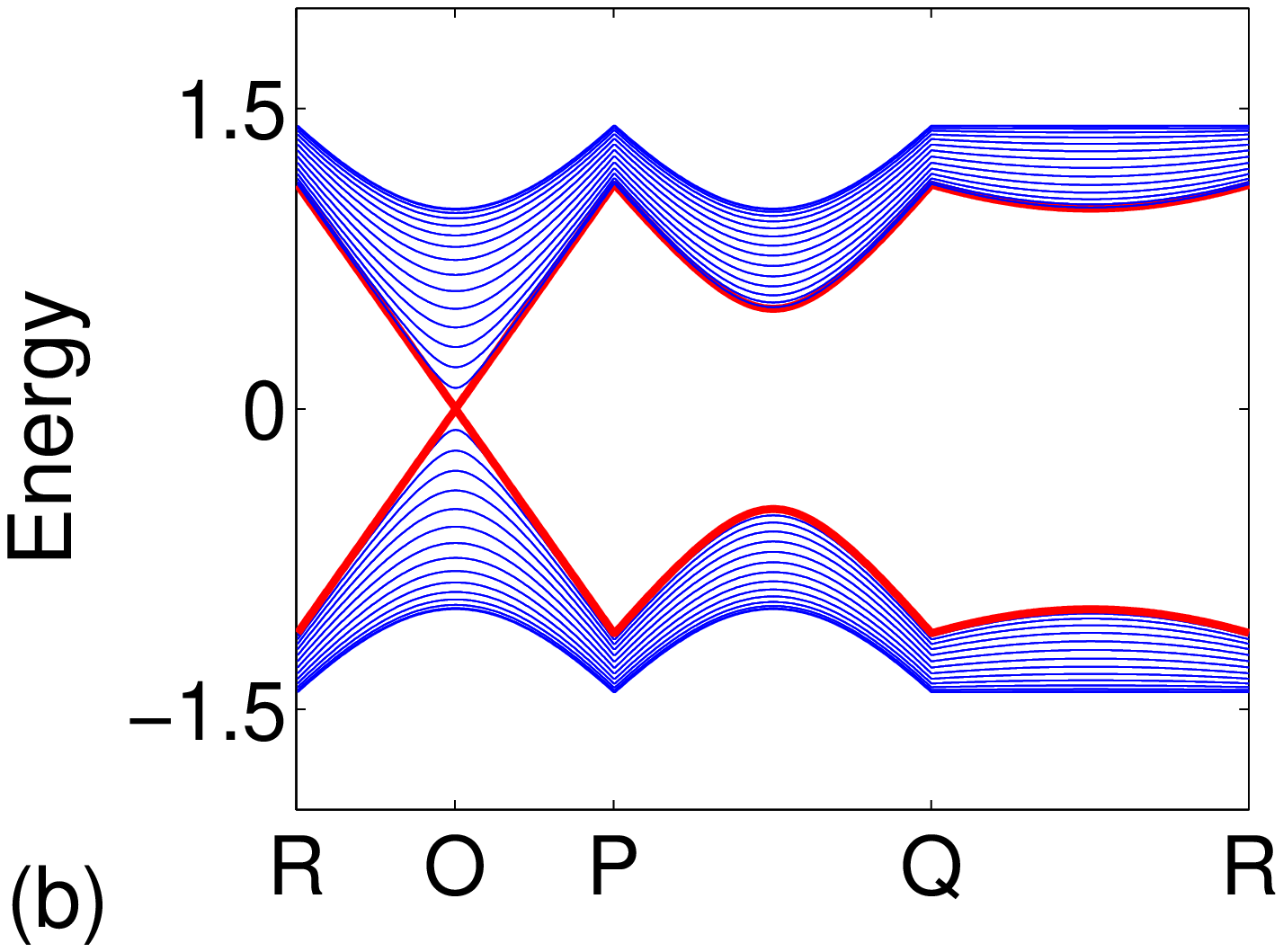} %
\includegraphics[ bb=89 236 514 553, width=0.3\textwidth, clip]{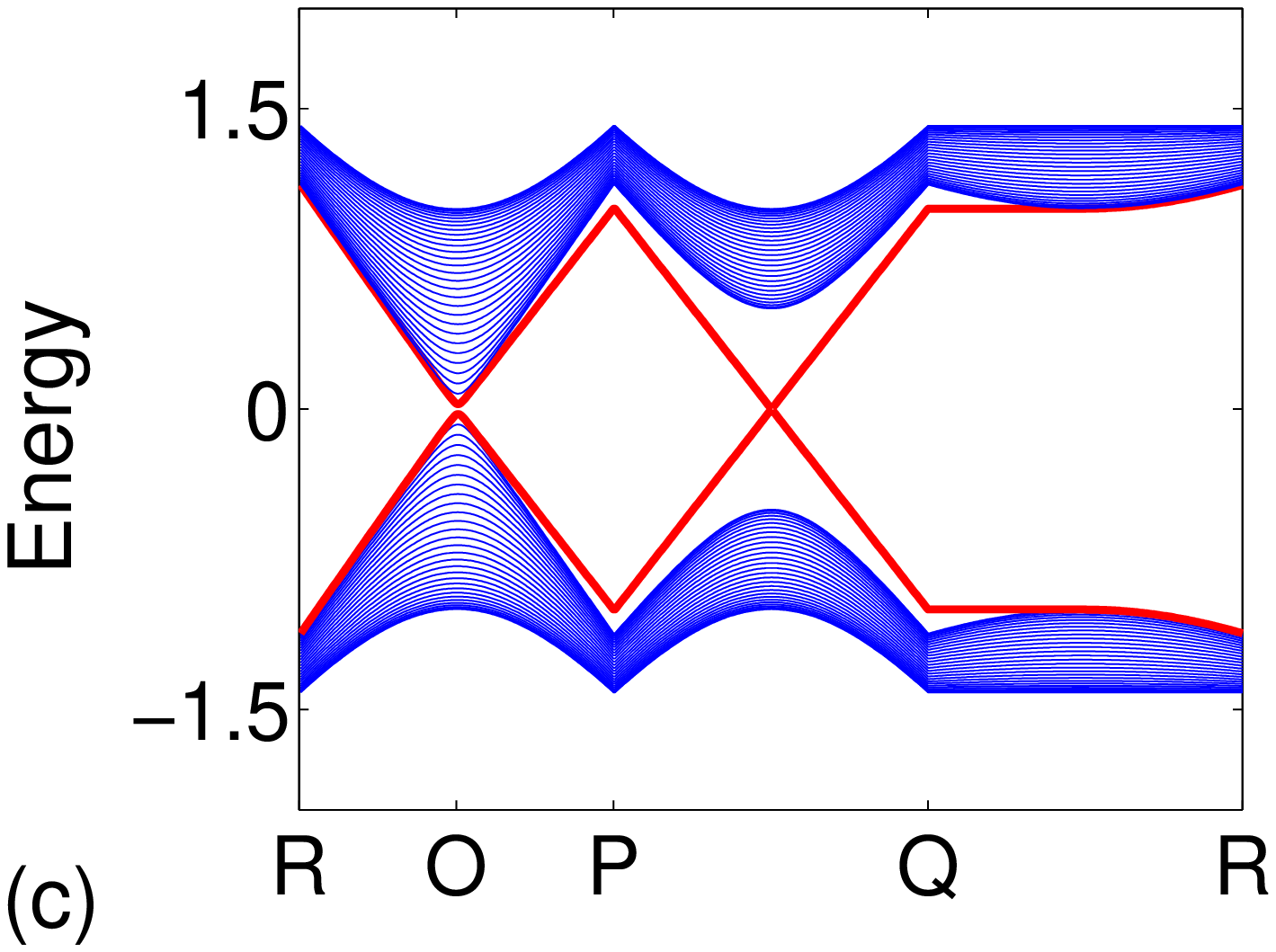}
\caption{(a) Schematics of $V\protect\delta $ plane. Energy spectrum for the
system with parameters at the sides of triangle PQROP for (b) periodic
boundary condition and (c) open boundary condition. The spectrum is
independent of the imbalance factor $\protect\lambda $. It indicates that
two bands touch at point $(0,0)$. There are two mid-gap states for (a) with $%
\protect\delta >0$, which are shown to be edge states with identical wave
functions for the standard SSH chain with $\protect\lambda =1$ and $V=0$.}
\label{fig1}
\end{figure*}

The non-Hermiticy arises from asymmetry factor $\lambda $, which\ has been
proposed to be realized in experiment\textbf{\ }\cite{S}. We note that the
Hamiltonian preserves time-reversal $(\mathcal{T})$\ symmetry. It has been
shown that such type of system has entirely full real spectrum \cite{ZXZ1}.
We introduce the Fourier transformations in two sub-lattices%
\begin{equation}
a_{l}=\frac{1}{\sqrt{N}}\sum_{k}e^{ikj}\left\{
\begin{array}{cc}
\beta _{k}, & l=2j \\
\alpha _{k}, & l=2j-1%
\end{array}%
\right. ,
\end{equation}%
where $j=1,2,...,N$, $k=2m\pi /N$, $m=0,1,2,...,N-1$. Spinless fermionic
operators in $k$ space $\alpha _{k},$\ $\beta _{k}$ are%
\begin{equation}
\left\{
\begin{array}{cc}
\beta _{k}=\frac{1}{\sqrt{N}}\sum\limits_{j}e^{-ikj}a_{l}, & l=2j \\
\alpha _{k}=\frac{1}{\sqrt{N}}\sum\limits_{j}e^{-ikj}a_{l}, & l=2j-1%
\end{array}%
\right. .
\end{equation}%
This transformation block diagonalizes the Hamiltonian due to its
translational symmetry, i.e.,%
\begin{equation}
H=\sum_{k\in \lbrack 0,2\pi )}H_{k}=\sum_{k\in \lbrack 0,2\pi )}\psi
_{k}^{\dagger }h_{k}\psi _{k},
\end{equation}%
satisfying $\left[ H_{k},H_{k^{\prime }}\right] =0$. Here\ $H$ is rewritten
in the Nambu representation with the basis%
\begin{equation}
\psi _{k}=\left(
\begin{array}{c}
\alpha _{k} \\
\beta _{k}%
\end{array}%
\right) ,
\end{equation}%
and $h_{k}$ is a $2\times 2$\ matrix%
\begin{equation}
h_{k}=\left(
\begin{array}{cc}
V & \lambda \gamma _{-k} \\
\lambda ^{-1}\gamma _{k} & -V%
\end{array}%
\right) ,  \label{h_k}
\end{equation}%
where $\gamma _{k}=(\gamma _{-k})^{\dagger }$ $=\frac{1}{2}[\left( 1-\delta
\right) +\left( 1+\delta \right) e^{ik}]$. The eigenvalues of $h_{k}$ are $%
|\varphi _{\rho }^{k}\rangle $ $\left( \rho =\pm \right) $ with eigenvalues%
\begin{equation}
\varepsilon _{\rho }^{k}=\rho \sqrt{|\gamma _{k}|^{2}+V^{2}}.
\label{spectrum}
\end{equation}%
In the case of nonzero $\lambda $, we have all real eigenvalue. Let us
denote the eigenvectors of a non-Hermitian Hamiltonian as%
\begin{eqnarray}
h_{k}|\varphi _{\rho }^{k}\rangle &=&\varepsilon _{\rho }^{k}|\varphi _{\rho
}^{k}\rangle , \\
h_{k}^{\dagger }|\eta _{\rho }^{k}\rangle &=&\varepsilon _{\rho }^{k}|\eta
_{\rho }^{k}\rangle ,  \notag
\end{eqnarray}%
where the explicit expression of $|\varphi _{\rho }^{k}\rangle $ and $|\eta
_{\rho }^{k}\rangle $ is%
\begin{equation}
|\varphi _{\rho }^{k}\rangle =\frac{1}{\Omega _{\rho }}\left(
\begin{array}{c}
V+\varepsilon _{\rho }^{k} \\
\frac{\gamma _{k}}{\lambda }%
\end{array}%
\right) ,
\end{equation}%
\begin{equation}
\left\vert \eta _{\rho }^{k}\right\rangle =\frac{1}{\Omega _{\rho }}\left(
\begin{array}{c}
V+\varepsilon _{\rho }^{k} \\
\lambda \gamma _{k}%
\end{array}%
\right) ,
\end{equation}%
where the normalization factors $\Omega _{\rho }=\sqrt{\left( V+\varepsilon
_{\rho }^{k}\right) ^{2}+|\gamma _{k}|^{2}}$. It is ready to check that
biorthogonal bases $\left\{ |\varphi _{\rho }^{k}\rangle ,|\eta _{\rho
}^{k}\rangle \right\} $ $\left( \rho =\pm \right) $ obey the biorthogonal
and completeness conditions%
\begin{equation}
\langle \eta _{\rho ^{\prime }}^{k^{\prime }}|\varphi _{\rho }^{k}\rangle
=\delta _{kk^{\prime }}\delta _{\rho \rho ^{\prime }},\sum_{\rho ,k}|\varphi
_{\rho }^{k}\rangle \langle \varphi _{\rho }^{k}|=1.  \label{bior}
\end{equation}%
There are two\ Bloch bands from the eigenvalues of $h_{k}$, indexed by $\rho
=\pm $. The band touching points\ occur at $k_{c}$\ when%
\begin{equation}
\sqrt{|\gamma _{k_{c}}|^{2}+V^{2}}=0.
\end{equation}%
The solutions of above equation indicates the degeneracy point%
\begin{equation}
\delta =0,V=0,
\end{equation}%
at $k_{c}=\pi $. The energy band structure is illustrated Fig. \ref{fig1}.
We would like to stress that the origin is a degeneracy point rather than an
exceptional point, although the system is a non-Hermitian system. The
exceptional point appears only at the case with $\lambda =0$ and $\infty $,
which is beyond our interest of this paper.

\section{Geometric quantities}

\label{Geometric quantities}

The topological property of the degeneracy point for a Hermitian RM model is
well established. It can be regarded as a monopole in the parameter space
\cite{Xiao}. We will investigate what happens when the Hamiltonian is
non-Hermitian. Before we focus on the concrete model, we first present a
general formalism for a non-Hermitian system. In this section, we will
develop a parallel theory for non-Hermitian case within the unbroken
symmetry region. Without loss of generality, we neglect the detail of the
model and consider a non-Hermitian spin-$1/2$\ system in an external
magnetic field, which can be described by the following Hamiltonian%
\begin{equation}
h(k,q)=\mathbf{B}(k,q)\cdot \mathbf{\sigma },
\end{equation}%
where $\mathbf{B}(k,q)=\left( B_{x},B_{y},B_{z},\right) $\ is a complex
magnetic field ($B_{x},B_{y}$\ are complex, and $B_{z}$\ is real) and $%
\sigma =\left( \sigma _{x},\sigma _{y},\sigma _{z}\right) $\ is Pauli
matrices%
\begin{equation}
\sigma _{x}=\left(
\begin{array}{cc}
0 & 1 \\
1 & 0%
\end{array}%
\right) ,\sigma _{y}=\left(
\begin{array}{cc}
0 & -i \\
i & 0%
\end{array}%
\right) ,\sigma _{z}=\left(
\begin{array}{cc}
1 & 0 \\
0 & -1%
\end{array}%
\right) .
\end{equation}%
The Hamiltonian is periodic with real $q$\ and $k$, i.e., $%
h(k,q)=h(k+k_{0},q)$ $=h(k,q+q_{0})$. Taking%
\begin{equation}
\left\{
\begin{array}{c}
B_{x}=B\cos \phi \sin \theta \\
B_{y}=B\sin \phi \sin \theta \\
B_{z}=B\cos \theta \\
\left\vert \mathbf{B}\right\vert =B=\sqrt{\left( B_{x}\right) ^{2}+\left(
B_{y}\right) ^{2}+\left( B_{z}\right) ^{2}}%
\end{array}%
\right. ,
\end{equation}%
with%
\begin{equation}
\tan \phi =B_{y}/B_{x},\cos \theta =B_{z}/B,
\end{equation}%
we rewrite the Hamiltonian as the form%
\begin{equation}
h=B\left(
\begin{array}{cc}
\cos \theta & \sin \theta e^{-i\phi } \\
\sin \theta e^{i\phi } & -\cos \theta%
\end{array}%
\right) .
\end{equation}%
We note that $\theta $\ is real, while $\phi $\ is complex. In parallel, for
the Hermitian conjugation counterpart%
\begin{equation}
h^{\dagger }=B\left(
\begin{array}{cc}
\cos \theta & \sin \theta e^{-i\phi ^{\ast }} \\
\sin \theta e^{i\phi ^{\ast }} & -\cos \theta%
\end{array}%
\right) ,
\end{equation}%
we have%
\begin{eqnarray}
h|\varphi _{\rho }\rangle &=&\rho B|\varphi _{\rho }\rangle , \\
h^{\dagger }|\eta _{\rho }\rangle &=&\rho B|\eta _{\rho }\rangle ,  \notag
\end{eqnarray}%
where $|\varphi _{\rho }\rangle $ and $|\eta _{\rho }\rangle $\textbf{\ }$%
\left( \rho =\pm \right) $\ is the instantaneous eigenvectors of $H$ and $%
H^{\dagger }$, respectively. Here the explicit expressions of these
eigenvectors are%
\begin{equation}
\left\vert \varphi _{+}\right\rangle =\left(
\begin{array}{c}
\cos \frac{\theta }{2} \\
\sin \frac{\theta }{2}e^{i\phi }%
\end{array}%
\right) ,\left\vert \varphi _{-}\right\rangle =\left(
\begin{array}{c}
-\sin \frac{\theta }{2} \\
\cos \frac{\theta }{2}e^{i\phi }%
\end{array}%
\right) ,
\end{equation}%
and%
\begin{equation}
\left\vert \eta _{+}\right\rangle =\left(
\begin{array}{c}
\cos \frac{\theta }{2} \\
\sin \frac{\theta }{2}e^{i\phi ^{\ast }}%
\end{array}%
\right) ,\left\vert \eta _{-}\right\rangle =\left(
\begin{array}{c}
-\sin \frac{\theta }{2} \\
\cos \frac{\theta }{2}e^{i\phi ^{\ast }}%
\end{array}%
\right) ,
\end{equation}%
which obey the biorthonormal relations%
\begin{equation}
\langle \eta _{\rho }|\varphi _{\rho ^{\prime }}\rangle =\langle \varphi
_{\rho ^{\prime }}|\eta _{\rho }\rangle =\delta _{\rho \rho ^{\prime }}.
\end{equation}%
We define the Berry connection $A_{\sigma }^{\rho }$\ $(\sigma =k,q)$\ and
the Berry curvature $\Omega _{kq}^{\rho }$\ in the context of biorthonormal
inner product%
\begin{equation}
A_{\sigma }^{\rho }=-i\langle \eta _{\rho }|\nabla _{\sigma }|\varphi _{\rho
}\rangle ,\Omega _{kq}^{\rho }=\partial _{k}A_{q}^{\rho }-\partial
_{q}A_{k}^{\rho },  \label{A and Omega}
\end{equation}%
Similar to the Hermitian case, $A_{\sigma }^{\rho }$\ is gauge dependent
while $\Omega _{k\rho }^{\rho }$\ is not. Accordingly Chern number can be
defined as%
\begin{equation}
c_{\rho }=\frac{1}{2\pi }\int_{k}^{k+k_{0}}\int_{q}^{q+q_{0}}\Omega
_{kq}^{\rho }dkdq.  \label{chern1}
\end{equation}%
For well defined $A_{k}^{\rho }$\ and $A_{q}^{\rho }$, the Chern number can
be rewritten as the form%
\begin{equation}
c_{\rho }=\frac{1}{2\pi }\oint_{\partial D}\mathcal{\mathbf{A}}^{\rho }\cdot
d\mathbf{r,}  \label{chern2}
\end{equation}%
where\textbf{\ }$\mathcal{\mathbf{A}}^{\rho }=A_{k}^{\rho }\hat{k}%
+A_{q}^{\rho }\hat{q}$\textbf{, }$r=k\hat{k}+q\hat{q}$\textbf{. }Here $\hat{q%
}$\ and $\hat{k}$\ are unit vectors, and $\partial D$\ (boundary of the
domain $D$, which covers the square of $k_{0}\times q_{0}$) is the path of
the integral.\textbf{\ }We concern whether $c_{\rho }$\ is still quantized
for real $\left\vert \mathbf{B}\right\vert $\ in the non-Hermitian system as
a topological invariant. We will investigate these issues based on a
concrete model.

\section{Chern number}

\label{Chern number}

In this section, we will calculate the explicit expressions of the geometric
quantities for the non-Hermitian RM model. The Bloch Hamiltonian is periodic
through the periodic functions $V(q)=V(q+2\pi )$ and $\delta (q)=\delta
(q+2\pi )$. When $q$ sweeps over a period, the system experiences a loop in
the $V\delta $\ plane.\ For this concrete model, we have the explicit form
of magnetic field%
\begin{equation}
\left\{
\begin{array}{c}
B_{x}=\frac{1}{2}\left( \lambda \gamma _{-k}+\lambda ^{-1}\gamma _{k}\right)
, \\
B_{y}=\frac{i}{2}\left( \lambda \gamma _{-k}-\lambda ^{-1}\gamma _{k}\right)
, \\
B_{z}=V,%
\end{array}%
\right.
\end{equation}%
and%
\begin{equation}
B=\sqrt{|\gamma _{k}|^{2}+V^{2}}.
\end{equation}%
Obviously, field $\mathbf{B}$ is periodic vector, $\mathbf{B}(q,k)=\mathbf{B}%
(q+2\pi ,k)$ $=\mathbf{B}(q,k+2\pi )$. In the following we only focus on the
lower band, neglecting the band index for the geometric quantity. Direct
derivations show that%
\begin{equation}
\mathcal{\mathbf{A}}=-i\langle \eta ^{k}|\partial _{q}|\varphi ^{k}\rangle
\hat{q}-i\langle \eta ^{k}|\partial _{k}|\varphi ^{k}\rangle \hat{k}
\end{equation}%
with%
\begin{eqnarray}
\langle \eta |\partial _{\sigma }|\varphi \rangle &=&\frac{1}{2}[\frac{1}{2}%
\sin \theta \left( \partial _{\sigma }\theta \right) \left( 1-e^{2i\text{%
\textrm{Re}}\phi }\right)  \notag \\
&&+i\left( 1+\cos \theta \right) \left( \partial _{\sigma }\phi \right) e^{2i%
\text{\textrm{Re}}\phi }],
\end{eqnarray}%
where $\hat{q}$\ and $\hat{k}$ are unit vectors for parameters $\sigma =q$, $%
k$ respectively, and the angles are%
\begin{eqnarray}
\phi &=&\arctan [\frac{i\left( \lambda \gamma _{-k}-\lambda ^{-1}\gamma
_{k}\right) }{\left( \lambda \gamma _{-k}+\lambda ^{-1}\gamma _{k}\right) }],
\label{phi} \\
\theta &=&\arccos (\frac{V}{\sqrt{|\gamma _{k}|^{2}+V^{2}}}).  \label{theta}
\end{eqnarray}%
We note that $\theta $\ is real, while $\phi $\ is complex. In contrast to a
Hermitian case, Berry connection $\mathcal{\mathbf{A}}$ is complex as
expected, which accords with the violation of the conservation of Dirac
probability. Apparently, the Berry curvature can be directly obtained by the
definition in Eq. (\ref{A and Omega}), and then the Chern number. However,
it is impossible to get all the Berry curvatures for a given loop through a
single expression of $\mathcal{\mathbf{A}}$, since the wave functions $%
\left\vert \varphi \right\rangle $ and $\left\vert \eta \right\rangle $\ are
not smooth and single valued everywhere.

For instance, when taking $\theta =0$, (or $B_{y}=B_{x}=0,$ and $B_{z}>0$)
we have%
\begin{equation}
\left\vert \varphi \right\rangle =\left(
\begin{array}{c}
0 \\
e^{i\phi }%
\end{array}%
\right) ,\left\vert \eta \right\rangle =\left(
\begin{array}{c}
0 \\
e^{i\phi ^{\ast }}%
\end{array}%
\right) .
\end{equation}%
where $\varphi $ is not well-defined, or indefinite. So $\left\vert \varphi
\right\rangle $ and $\left\vert \eta \right\rangle $ have singularity at $%
\mathbf{B}=\left( 0,0,B_{z}\right) $ with $B_{z}>0$. We can choose another
gauge by multiplying $\left\vert \varphi \right\rangle $ and $\left\vert
\eta \right\rangle $ by $e^{-i\mathrm{Re}\phi }$. Then the wave function is
smooth and single valued everywhere expect at the south pole, i.e., $\varphi
$ is not well-defined, or indefinite at $\theta =\pi $. For a given loop,
one should choose one or two specific gauges, or expressions of $\left\vert
\varphi \right\rangle $ and $\left\vert \eta \right\rangle $\ in order to
calculate the Chern number. If a set of proper chosen wave functions $%
\left\{ \left\vert \varphi \right\rangle ,\left\vert \eta \right\rangle
\right\} $\ are well-defined for a given loop, the Chern number is zero. If
a loop requires two different gauges, the Chern number is nonzero.
Obviously, the difference between the Berry connections $\mathcal{\mathbf{A}}
$\ in two gauges is $\nabla _{\mathbf{R}}\mathrm{Re}\phi $,\ and thus Chern
number are definitely real.

We choose four typical loops, involving: (a) quadrants I and II $(V>0)$; (b)
quadrants III and IV $(V<0)$; (c) one of four quadrants, or quadrants I and
IV $(\delta >0)$, or quadrants II and III $(\delta <0)$, and (d) quadrants
I, II, III, and IV (enclosing the origin). For the cases (a-c), one can take
the gauge as%
\begin{eqnarray}
\text{(a)} &:&\left\vert \varphi \right\rangle ,\left\vert \eta
\right\rangle ;  \notag \\
\text{(b)} &:&e^{i\mathrm{Re}\phi }\left\vert \varphi \right\rangle ,e^{i%
\mathrm{Re}\phi }\left\vert \eta \right\rangle ;  \notag \\
\text{(c)} &:&\left\vert \varphi \right\rangle ,\left\vert \eta
\right\rangle ;  \notag \\
&&\text{or }e^{i\mathrm{Re}\phi }\left\vert \varphi \right\rangle ,e^{i%
\mathrm{Re}\phi }\left\vert \eta \right\rangle ,
\end{eqnarray}%
while one need two gauges for (d) as%
\begin{eqnarray}
\text{(dI)} &:&\left\vert \varphi \right\rangle ,\left\vert \eta
\right\rangle ,\text{ }\left( V>0\right) ;  \notag \\
\text{(dII)} &:&e^{i\mathrm{Re}\phi }\left\vert \varphi \right\rangle ,e^{i%
\mathrm{Re}\phi }\left\vert \eta \right\rangle ,\text{ }\left( V<0\right) .
\end{eqnarray}%
We conclude from this analysis that the Chern number is nonzero when the
loop encloses the origin. Explicitly, with the Eq. (\ref{chern2}), we have%
\begin{eqnarray}
c &=&\frac{1}{2\pi }(\oint_{\partial D_{\text{I}}}\mathcal{\mathbf{A}}_{%
\text{I}}+\oint_{\partial D_{\text{II}}}\mathcal{\mathbf{A}}_{\text{II}%
})\cdot d\mathbf{r}  \notag \\
&=&\frac{1}{2\pi }\oint_{\partial D_{\text{I}}}\left( \mathcal{\mathbf{A}}_{%
\text{I}}-\mathcal{\mathbf{A}}_{\text{II}}\right) \cdot d\mathbf{r}  \notag
\\
&\neq &0,  \label{chern3}
\end{eqnarray}%
where the subindex of $\mathcal{A}_{\text{I}}$\ and $\mathcal{A}_{\text{II}}$%
\textbf{\ }distinguish $\left( \left\vert \varphi \right\rangle ,\left\vert
\eta \right\rangle \right) $\ from $\left( e^{i\mathrm{Re}\phi }\left\vert
\varphi \right\rangle ,e^{i\mathrm{Re}\phi }\left\vert \eta \right\rangle
\right) $\ respectively, when we define the Berry connection.

To demonstrate the result, we consider a loop in the form%
\begin{equation}
\left\{
\begin{array}{c}
\delta =\delta _{c}+r\cos q \\
V=V_{c}+r\sin q%
\end{array}%
\right.  \label{cir loop}
\end{equation}%
which is a circle in $V\delta $ plane with radius $r$, centered at $\left(
\delta _{c},V_{c}\right) $. The Chern number can be obtained as

\begin{equation}
c=\left\{
\begin{array}{cc}
0, & \delta _{c}^{2}+V_{c}^{2}>r^{2} \\
1, & \delta _{c}^{2}+V_{c}^{2}<r^{2}%
\end{array}%
\right. ,  \label{c cir}
\end{equation}%
i.e., Chern number is zero if the loop does not encircle the degeneracy
point, while is nonzero if encircle the degeneracy point. In addition, if we
take $q\rightarrow -q$, we will get $c\rightarrow -c$, which means that one
gets opposite Chern number for the same loop but with opposite direction. We
schematically illustrate this loop in Fig. \ref{fig3}(a).

It shows that the degeneracy point still has topological feature in a
non-Hermitian system even though the Berry connection may be complex. The
underlying mechanism of the quantization of the Chern number for a
non-Hermitian system is that the single-valuedness of the wave function is
always true no matter the system is Hermitian or not.

\section{Edge-mode operators}

\label{Edge-mode operators}

In a Hermitian SSH model, the degenerate zero modes take the role of
topological invariant for opened chain. In this section, we investigate the
similar feature for non-Hermitian RM model. Considering the RM model with an
open boundary condition, the Hamiltonians read%
\begin{eqnarray}
&&H_{\mathrm{CH}}=H-M,  \notag \\
&&M=\kappa _{2N,1}a_{2N}^{\dagger }a_{1}+\kappa _{1,2N}a_{1}^{\dagger
}a_{2N},
\end{eqnarray}%
which represents the original system with broken the coupling across two
sites $(2N,1)$. We introduce a pair of edge-mode operators $\overline{A}_{%
\mathrm{L,R}}$ in the infinite $N$ limit, which are defined as%
\begin{eqnarray}
\overline{A}_{\text{L}} &=&\frac{1}{\sqrt{\Omega }}\sum\limits_{j=1}^{N}%
\left( \frac{\delta -1}{\delta +1}\right) ^{N-j}a_{2j}^{\dagger }, \\
\overline{A}_{\text{R}} &=&\frac{1}{\sqrt{\Omega }}\sum\limits_{j=1}^{N}%
\left( \frac{\delta -1}{\delta +1}\right) ^{j-1}a_{2j-1}^{\dagger },
\end{eqnarray}%
where $\Omega =\{1-\left[ \left( \delta -1\right) /\left( \delta +1\right) %
\right] ^{2N}\}/\{1-\left[ \left( \delta -1\right) /\left( \delta +1\right) %
\right] ^{2}\}$. It is easy to check that%
\begin{eqnarray}
\lbrack \overline{A}_{\text{L}},H_{\mathrm{CH}}] &=&V\overline{A}_{\text{L}},
\\
\lbrack \overline{A}_{\text{R}},H_{\mathrm{CH}}] &=&-V\overline{A}_{\text{R}%
},
\end{eqnarray}%
which ensues that
\begin{eqnarray}
H_{\mathrm{CH}}\overline{A}_{\text{L}}\left\vert \text{Vac}\right\rangle
&=&-V\overline{A}_{\text{L}}\left\vert \text{Vac}\right\rangle , \\
H_{\mathrm{CH}}\overline{A}_{\text{R}}\left\vert \text{Vac}\right\rangle &=&V%
\overline{A}_{\text{R}}\left\vert \text{Vac}\right\rangle , \\
H_{\mathrm{CH}}\overline{A}_{\text{L}}\overline{A}_{\text{R}}\left\vert
\text{Vac}\right\rangle &=&0\times \overline{A}_{\text{L}}\overline{A}_{%
\text{R}}\left\vert \text{Vac}\right\rangle .
\end{eqnarray}%
Here $\left\vert \text{Vac}\right\rangle $ is the vacuum state of fermion
operator, i.e., $a_{l}\left\vert \text{Vac}\right\rangle =0$. Obviously,
states $\overline{A}_{\text{L,R}}\left\vert \text{Vac}\right\rangle $\ and $%
\overline{A}_{\text{L}}\overline{A}_{\text{R}}\left\vert \text{Vac}%
\right\rangle $\ are the eigenstates of $H_{\mathrm{CH}}$\ with eigen
energies $\mp V$\ and $0$, respectively.

In parallel, we can also define the biorthogonal conjugation operators%
\begin{eqnarray}
A_{\text{L}} &=&\frac{1}{\sqrt{\Omega }}\sum\limits_{j=1}^{N}\left( \frac{%
\delta -1}{\delta +1}\right) ^{N-j}a_{2j}, \\
A_{\text{R}} &=&\frac{1}{\sqrt{\Omega }}\sum\limits_{j=1}^{N}\left( \frac{%
\delta -1}{\delta +1}\right) ^{j-1}a_{2j-1},
\end{eqnarray}%
which satisfy the canonical commutation relations
\begin{equation}
\{A_{\mu },\overline{A}_{\nu }\}=\delta _{\mu \nu },\{A_{\mu },A_{\nu }\}=\{%
\overline{A}_{\mu },\overline{A}_{\nu }\}=0,
\end{equation}%
with the indices $\mu $, $\nu =$ L, R. Similarly, we have%
\begin{equation}
\lbrack A_{\text{L}}^{\dag },H_{\mathrm{CH}}^{\dag }]=VA_{\text{L}}^{\dag
},[A_{\text{R}}^{\dag },H_{\mathrm{CH}}^{\dag }]=-VA_{\text{R}}^{\dag },
\end{equation}%
which indicate that states $A_{\text{L,R}}^{\dag }\left\vert \text{Vac}%
\right\rangle $\ and $A_{\text{L}}^{\dag }A_{\text{R}}^{\dag }\left\vert
\text{Vac}\right\rangle $\ are the eigenstates of $H_{\mathrm{CH}}^{\dag }$\
with eigen energies $\mp V$\ and $0$, respectively.

\begin{figure}[tbp]
\includegraphics[ bb=97 226 504 563, width=0.4\textwidth, clip]{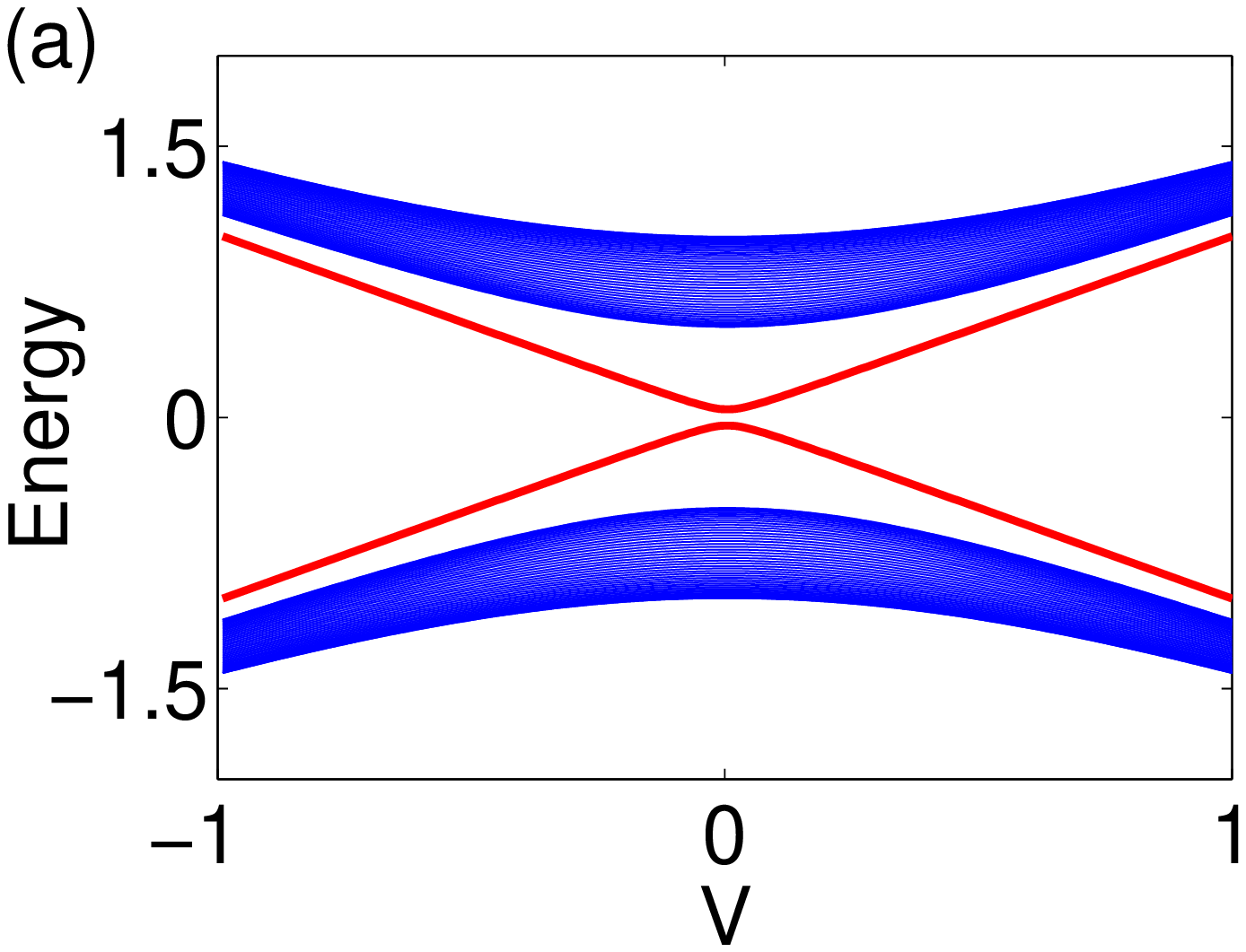}
\includegraphics[ bb=51 186 539 764, width=0.43\textwidth,
clip]{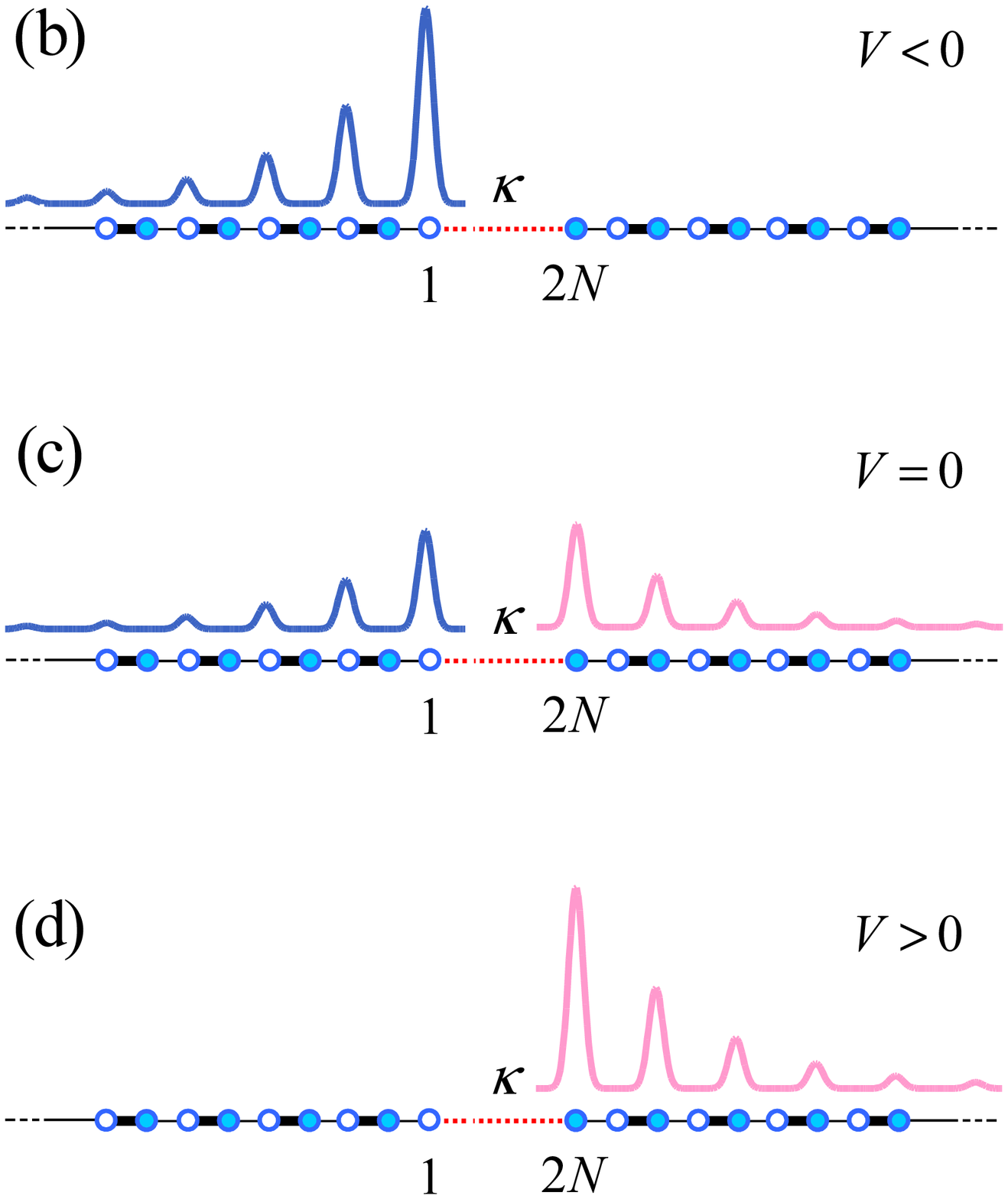}
\caption{(a) Plot of the spectrum of $H_{\protect\kappa }$ in Eq. (\protect
\ref{H_cappa}) with $N=50$, $\protect\delta =0.5,$ $\protect\kappa =0.05,$
and $\protect\lambda =1.5$. Comparing to that in Fig. \protect\ref{fig1}(c),
the level crossing becomes avoided level crossing around $V=0$, at which two
types of edge modes are hybridized. (b) and (d) are schematics for two edge
modes. (c) The superpositions of the two become mid-gap states when $V$
turns to zero. When varying $V$\ from $-1$\ to $1$\ adiabatically, (b) will
evolves to (c), then (d), resulting an adiabatic particle transport from the
leftmost of the chain to the rightmost.}
\label{fig2}
\end{figure}

\begin{figure}[tbp]
\includegraphics[ bb=33 63 477 790, width=0.5\textwidth, clip]{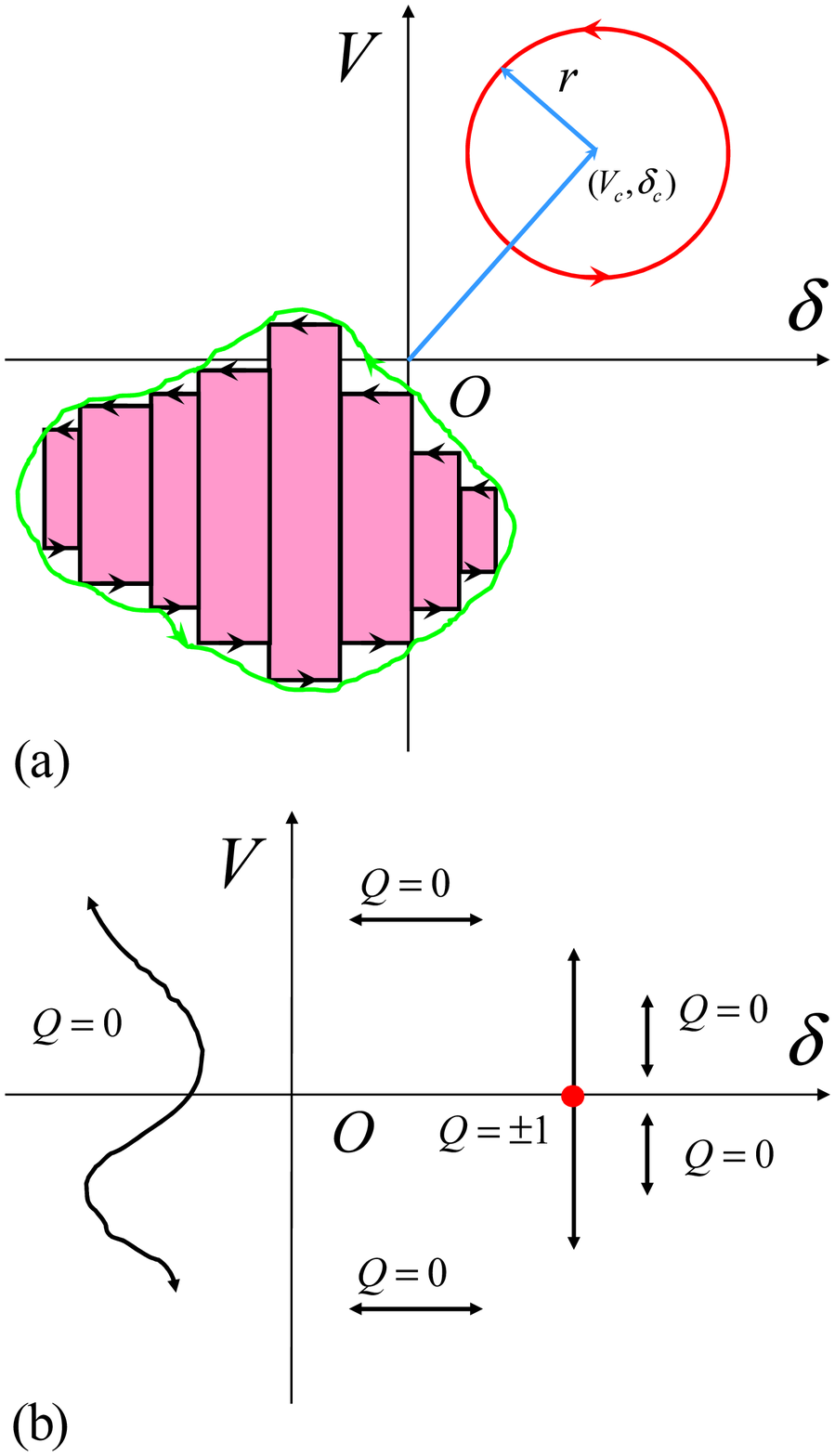}
\caption{(a) Schematics of a simple loop represented in Eq. (\protect\ref{c
cir}). For an arbitrary loop, it can be decomposed into many rectangular
loops. The Chern number of an arbitrary loop is the sum of all the Chern
numbers of the rectangles. (b) Six types of segments which can be used to
construct any kinds of loops. It turns out that only the vertical one
passing through the red dot has nonzero charge transport across the two ends
of the chain (see text and Fig. \protect\ref{fig2}(b)-(d)).}
\label{fig3}
\end{figure}

A surprising fact is that $A_{\mu }^{\dag }=\overline{A}_{\mu }$, which does
not hold true in general since $H_{\mathrm{CH}}\neq H_{\mathrm{CH}}^{\dag }$%
. It is due to the special eigenstates of a particular model. This feature
allows us to treat the edge modes in the framework of Hermitian regime. The
biorthonormal and Dirac probabilities of the edge modes are the same and can
be expressed as%
\begin{equation}
\mathcal{P}_{\mu }(l)=\left\langle \text{Vac}\right\vert A_{\mu
}a_{l}^{\dagger }a_{l}A_{\mu }^{\dag }\left\vert \text{Vac}\right\rangle .
\end{equation}%
or explicit form

\begin{equation}
\mathcal{P}_{\mathrm{L}}(l)=\left\{
\begin{array}{cc}
\frac{1}{\Omega }\left( \frac{\delta -1}{\delta +1}\right) ^{2N-l}, & l=2j
\\
0, & l=2j-1%
\end{array}%
\right. ,
\end{equation}%
and

\begin{equation}
\mathcal{P}_{\mathrm{R}}(l)=\left\{
\begin{array}{cc}
0, & l=2j \\
\frac{1}{\Omega }\left( \frac{\delta -1}{\delta +1}\right) ^{l-1}, & l=2j-1%
\end{array}%
\right. ,
\end{equation}%
which obey the normalization condition $\sum_{l}\mathcal{P}_{\mu }(l)=1$. We
note that the edge modes are independent of $\lambda $ and $V$, identical to
that for the standard SSH chain ($V=0$ and $\lambda =1$). In contrast to the
standard SSH chain, the eigenvalues of edged modes can be nonzero. The
profiles of the edge modes are illustrated schematically in Fig. \ref{fig2}.

Based on the above analysis, it turns out that the bulk system exhibits the
similar topological feature when taking $V=0$. In Hermitian systems, the
existence of edge modes is intimately related to the bulk topological
quantum numbers, which is referred as the bulk-edge correspondence relations
\cite{Thouless,Kane,Zhang,Lu}. We are interested in the generalization of
the bulk-edge correspondence to this non-Hermitian system. Previous works
show that when sufficiently weak non-Hermiticity is introduced to
topological insulator models, the edge modes can retain some of their
original characteristics \cite{Esaki,Hu}. Similarly, we can get the same
conclusion for the case with $V=0$, if we define the Zak phase in the
framework of biorthonormal inner product, demonstrating the bulk-edge
correspondence.

\begin{figure}[tbp]
\begin{minipage}{0.48\linewidth}
\centerline{\includegraphics[width=4.6cm]{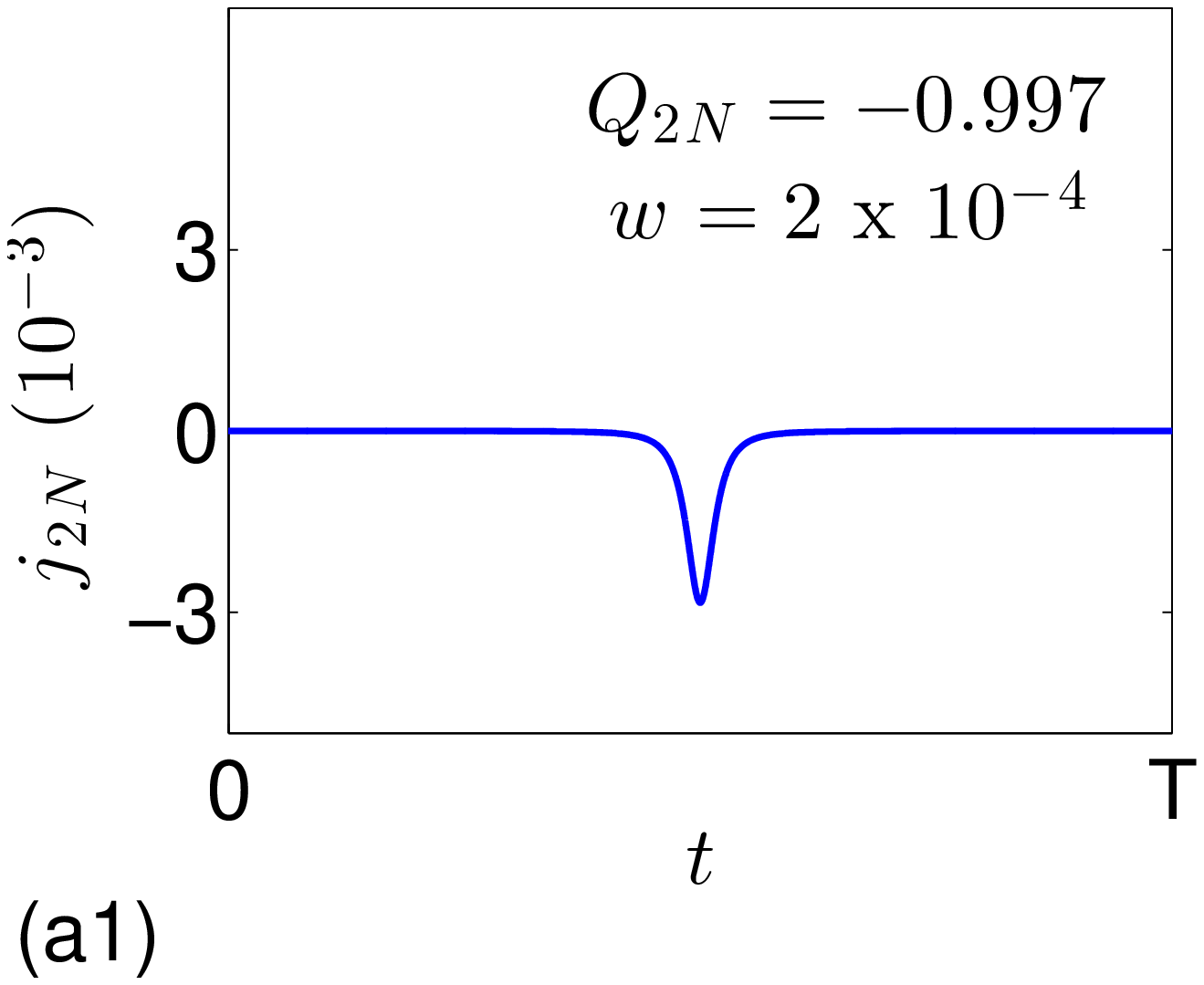}}
\end{minipage}
\begin{minipage}{0.48\linewidth}
\centerline{\includegraphics[width=4.6cm]{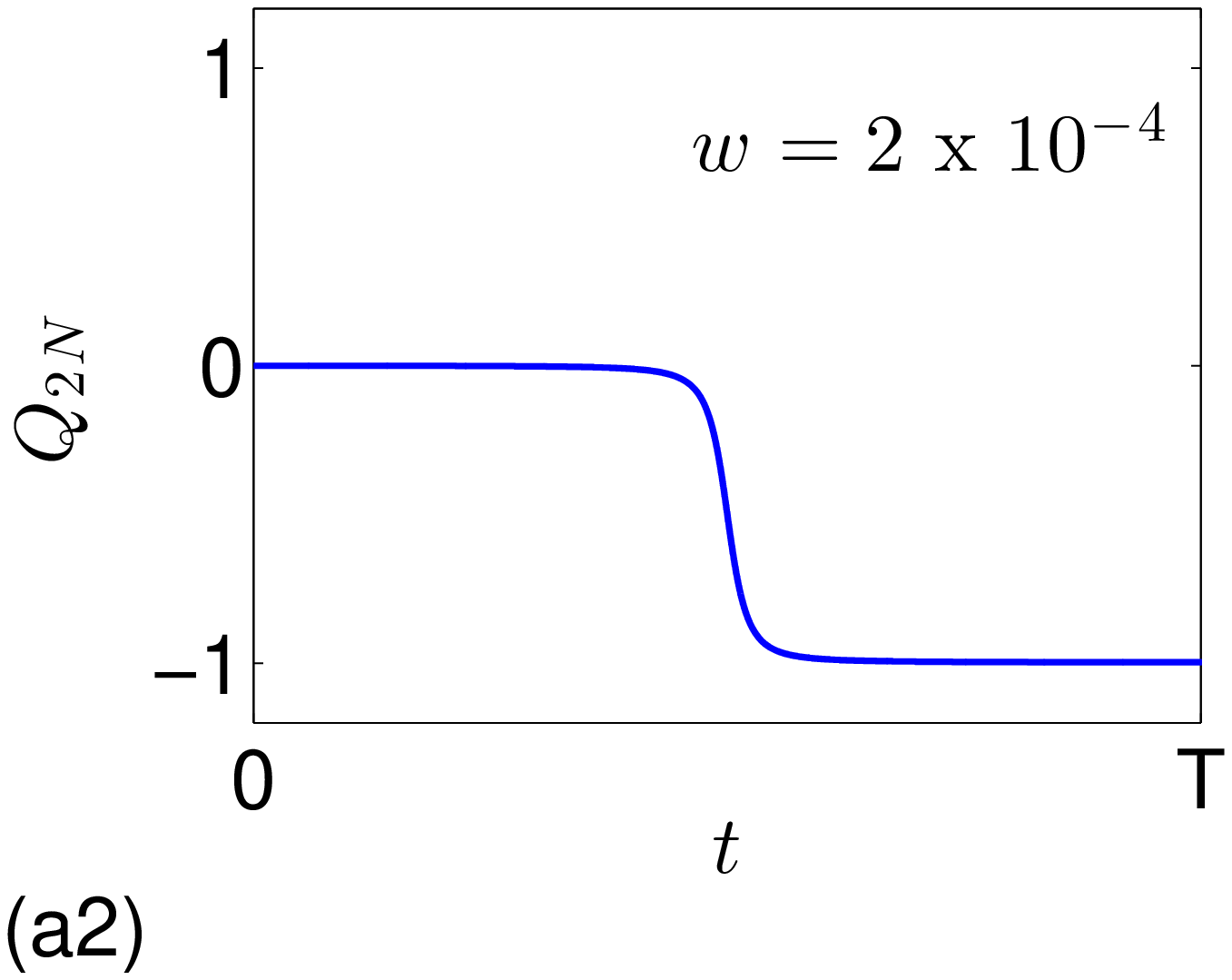}}
\end{minipage}
\begin{minipage}{0.48\linewidth}
\centerline{\includegraphics[width=4.6cm]{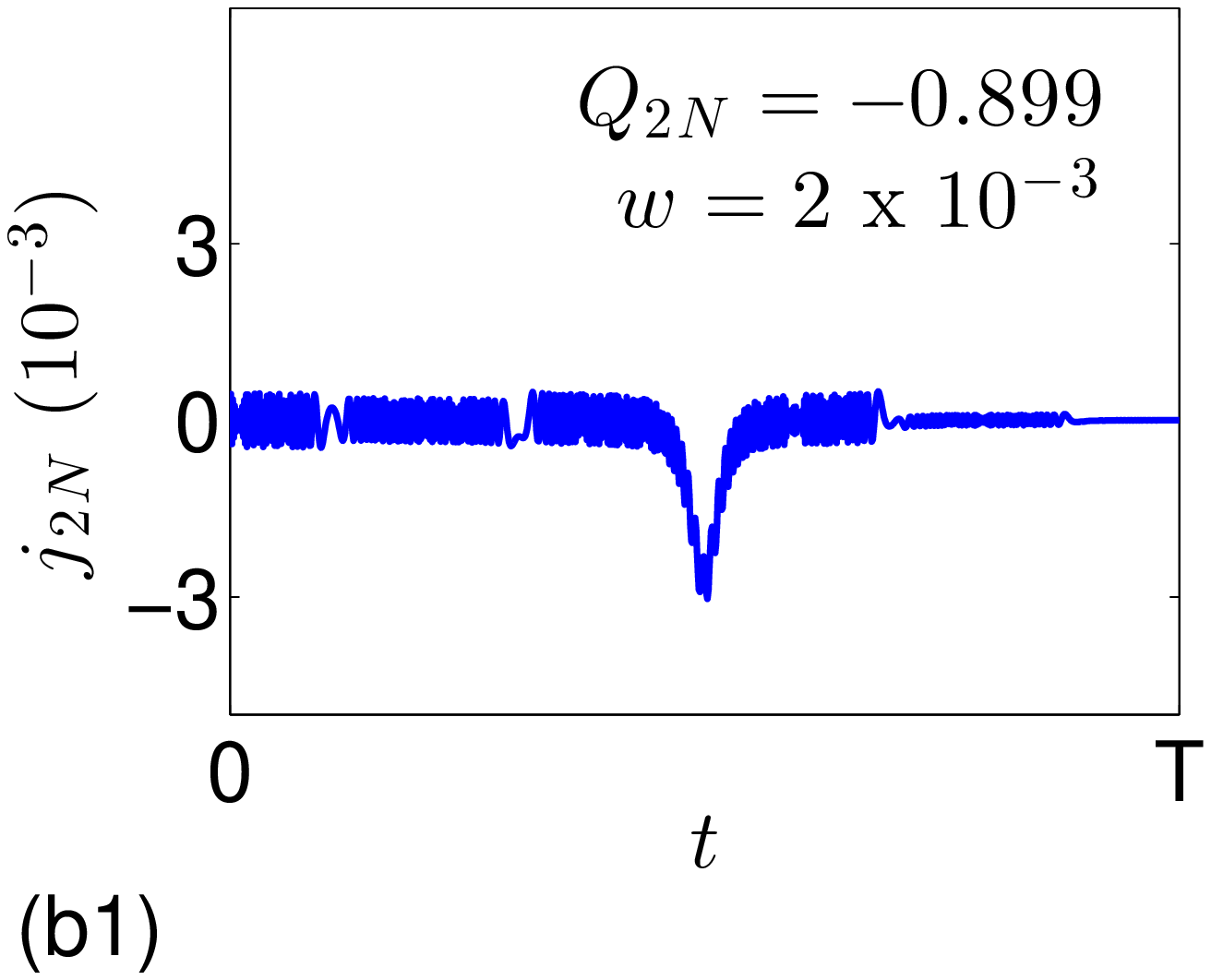}}
\end{minipage}
\begin{minipage}{0.48\linewidth}
\centerline{\includegraphics[width=4.6cm]{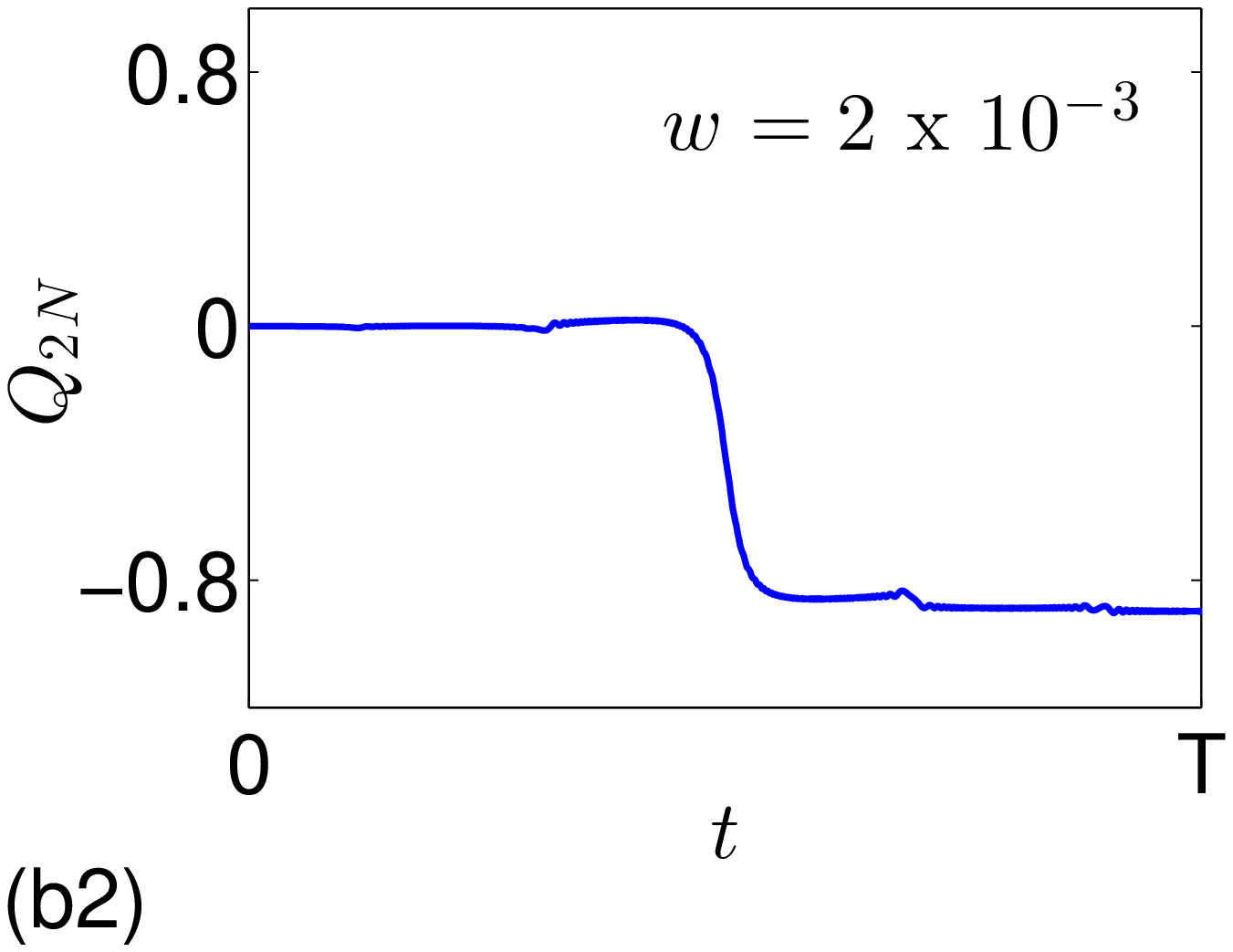}}
\end{minipage}
\begin{minipage}{0.48\linewidth}
\centerline{\includegraphics[width=4.6cm]{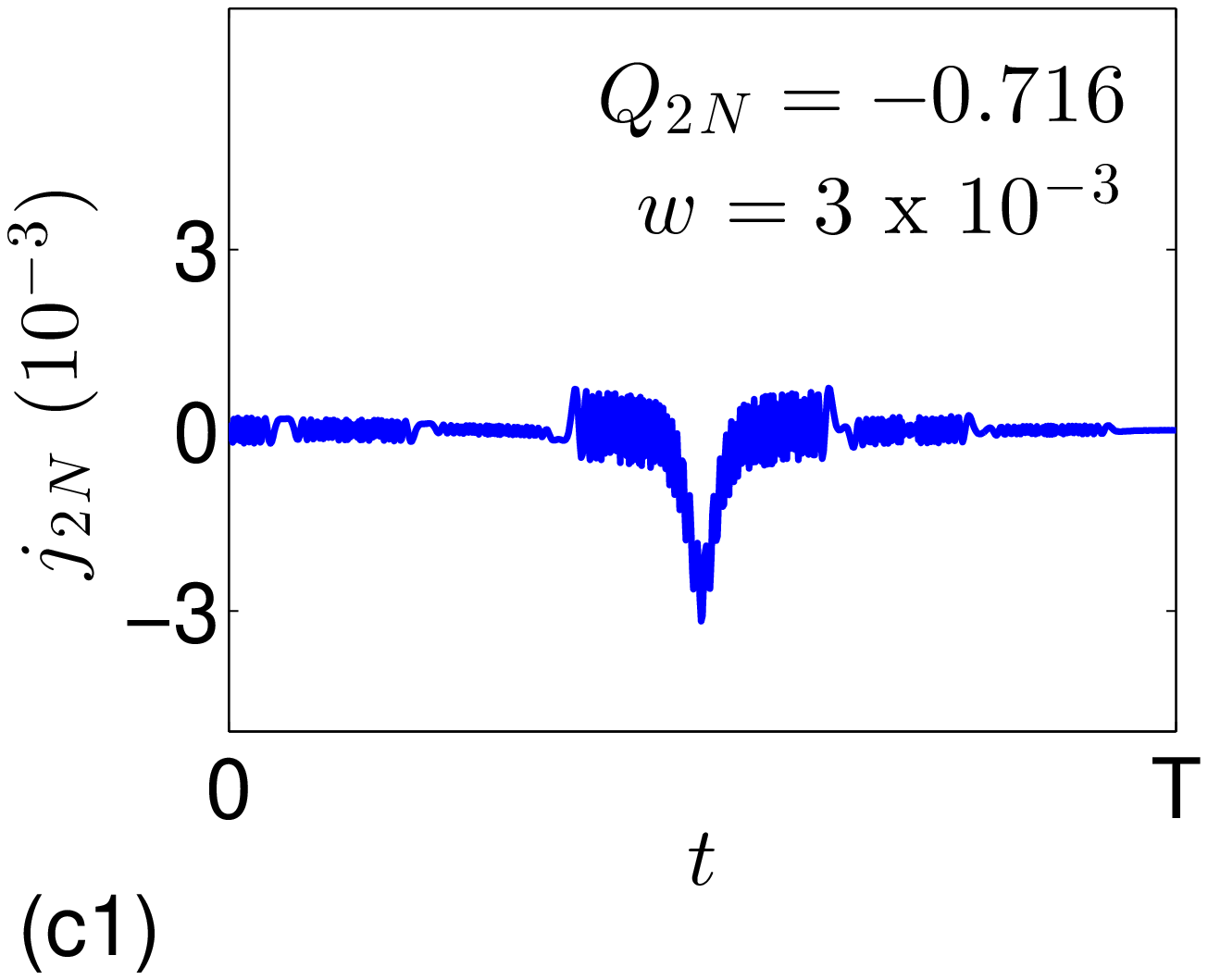}}
\end{minipage}
\begin{minipage}{0.48\linewidth}
\centerline{\includegraphics[width=4.6cm]{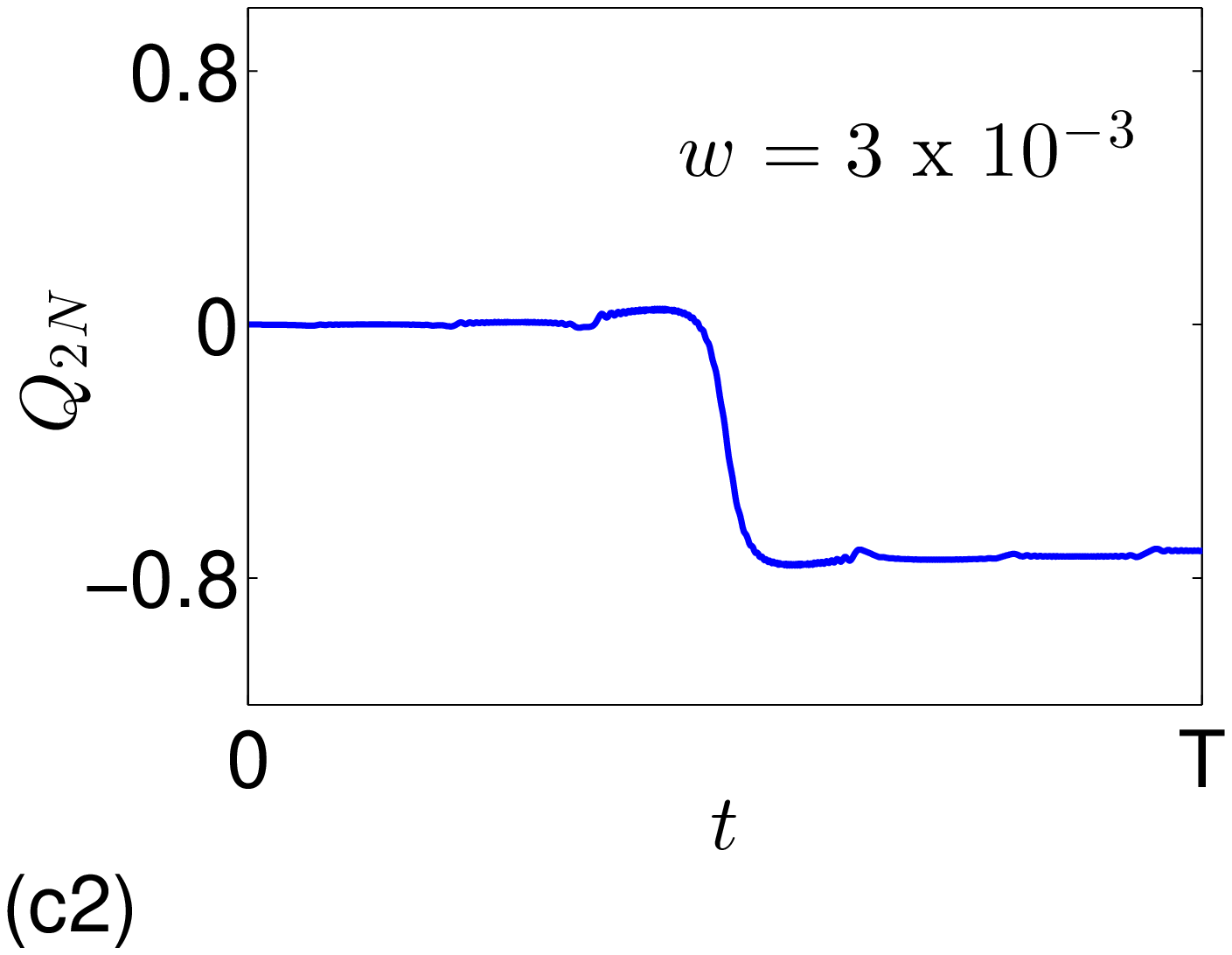}}
\end{minipage}
\begin{minipage}{0.48\linewidth}
\centerline{\includegraphics[width=4.6cm]{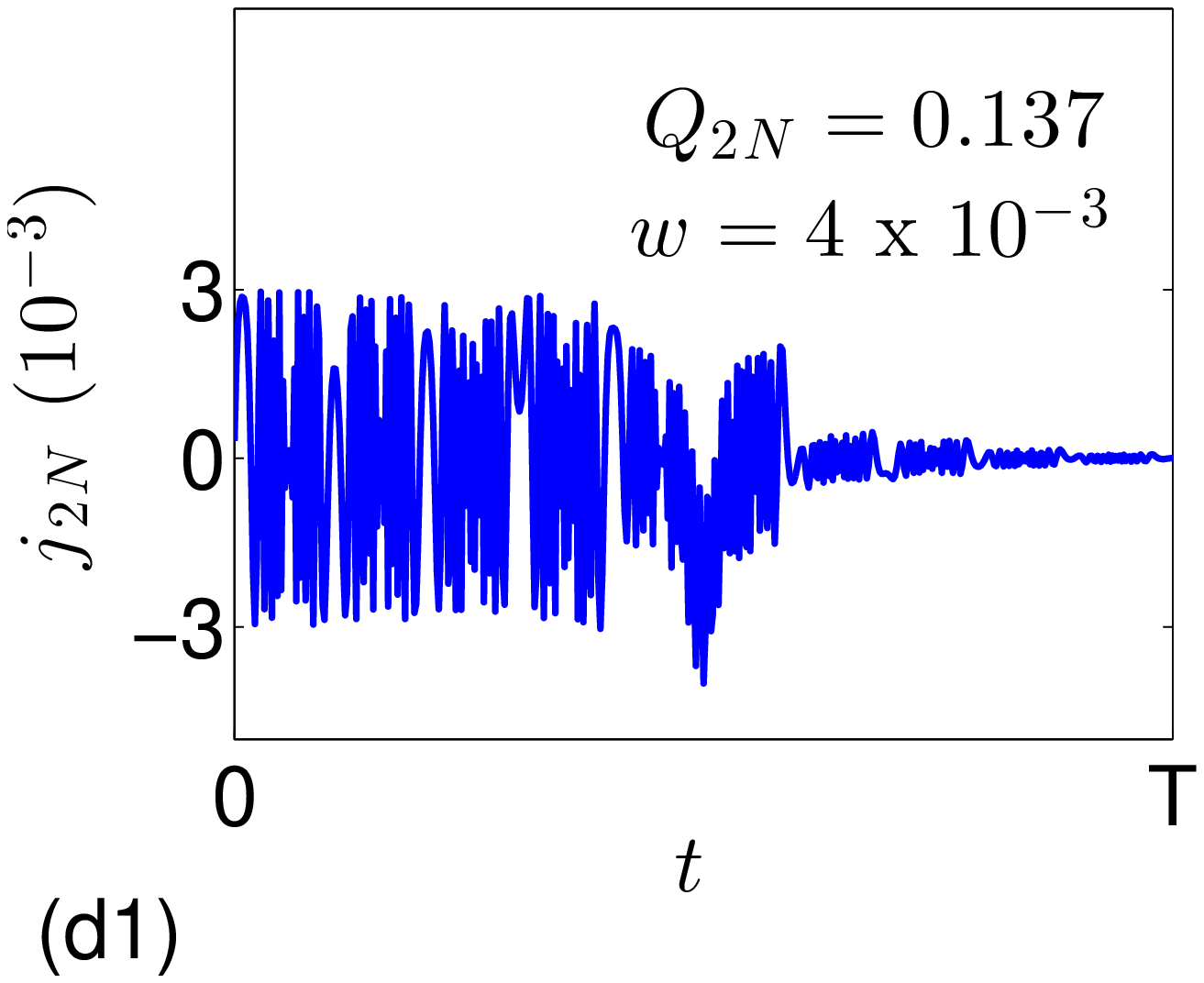}}
\end{minipage}
\begin{minipage}{0.48\linewidth}
\centerline{\includegraphics[width=4.6cm]{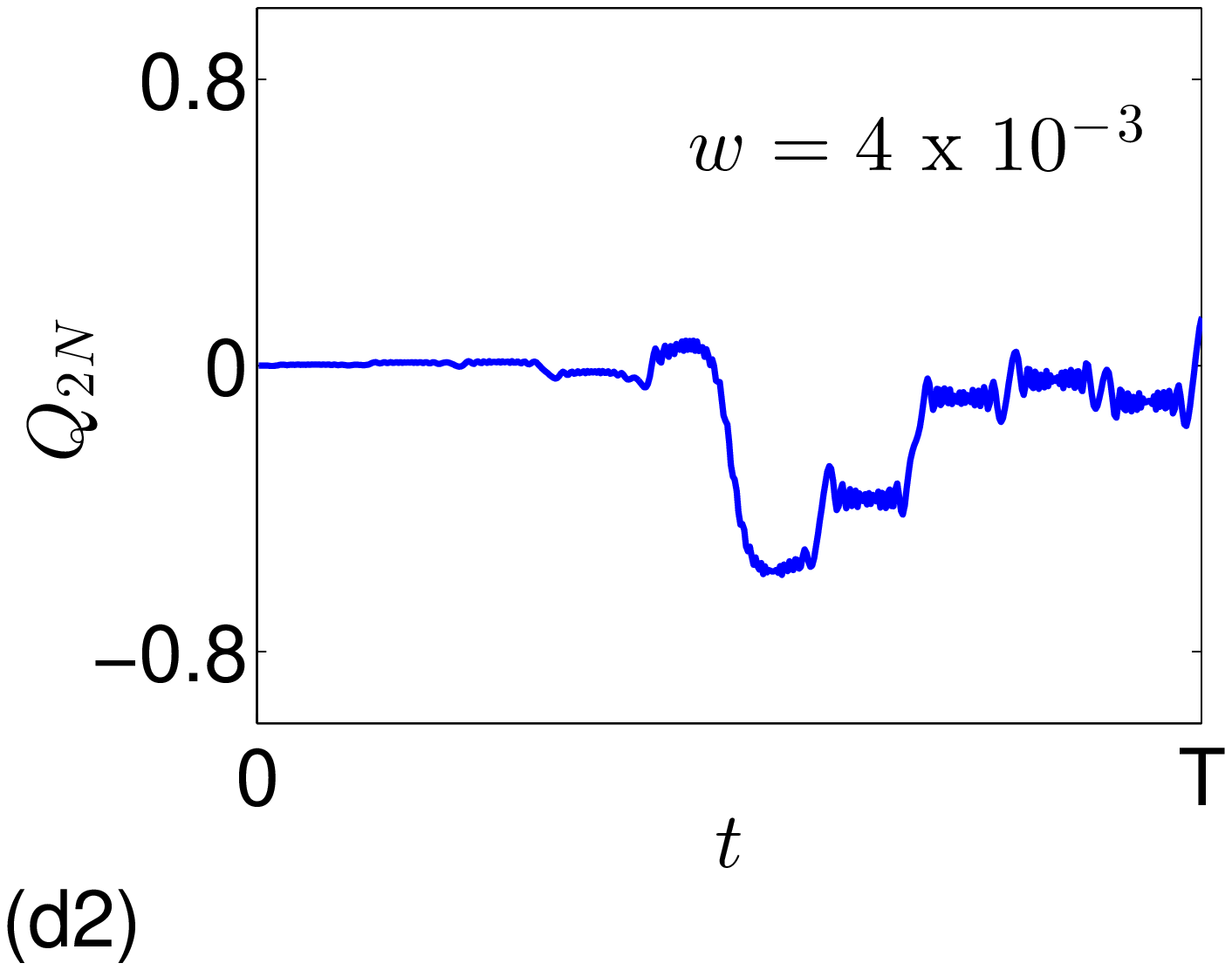}}
\end{minipage}
\caption{Numerical simulations of particle transport across two ends of the
chain current obtained by exact diagonalization, driven by the
time-dependent parameter $V=1-2\protect\omega t$, $t\in \left[ 0,T=\protect%
\omega ^{-1}\right]$. Plots of the current (a1-d1) and charge accumulation
as function of time (a2-d2) for several typical speed $\protect\omega $,
indicated in the figures. The total charge transfer during the interval $[0,%
\protect\omega ^{-1}]$\ is also indicated in each figure.\ We note that the
result accords with our prediction well when the speed is slow enough in
(a1, b1). As $\protect\omega $\ increases, the deviation becomes larger and
larger. In the case of (a4, b4), the accumulated charge becomes positive,
which is qualitatively different from the adiabatic process. The parameters
of the system are $N=10$, $\protect\delta =0.5$, $\protect\kappa =0.05$, and
$\protect\lambda =1.5$.}
\label{fig4}
\end{figure}

\section{Pumping charge}

\label{Pumping charge}

In a Hermitian system, the physical meaning of\ Chern number is well known.
In a Hermitian RM model, it has been shown that the adiabatic particle
transport over a time period takes the form of the Chern number\ and it is
quantized \cite{Xiao}. The pumped charge counts the net number of degenerate
point enclosed by the loop. For example, consider a time-dependent
Hamiltonian, which varies adiabatically along the circle in Eq. (\ref{cir
loop}). After a period of time, the particle transport for the half-filled
ground state equals to the Chern number of the loop in Eq. (\ref{c cir}). In
the above, we have shown that a non-Hermitian system can exhibit the similar
topological feature as that in a Hermitian system. The Chern number as the
topological invariant defined in the context of biorthonormal inner product.
In parallel, such a Chern number\ should have the similar physical meaning.

Actually, one can rewrite Eq. (\ref{chern2}) in the form%
\begin{eqnarray}
c &=&\oint (\frac{\partial \mathcal{Z}}{\partial \delta }d\delta +\frac{%
\partial \mathcal{Z}}{\partial V}dV)  \notag \\
&=&\int_{0}^{T}(\frac{\partial \mathcal{Z}}{\partial \delta }\dot{\delta}+%
\frac{\partial \mathcal{Z}}{\partial V}\dot{V})dt,
\end{eqnarray}%
where $\mathcal{Z}$\ is Zak phase%
\begin{equation}
\mathcal{Z}=\frac{i}{2\pi }\int_{0}^{2\pi }\langle \eta ^{k}|\partial
_{k}|\varphi ^{k}\rangle dk,
\end{equation}%
defined in the context of biorthonormal inner product. Here $V$\ and $\delta
$\ is periodic function of time $t$\ and the sub-index is neglected for
simplicity. Furthermore, we can find out the physical meaning of the Chern
number by the relation
\begin{equation}
c=\int_{0}^{T}\mathcal{J}(t)dt,
\end{equation}%
where%
\begin{equation}
\mathcal{J}=\frac{i}{2\pi }\int_{0}^{2\pi }[(\partial _{t}\langle \eta
^{k}|)\partial _{k}|\varphi ^{k}\rangle -(\partial _{k}\langle \eta
^{k}|)\partial _{t}|\varphi ^{k}\rangle ]dk
\end{equation}%
is the adiabatic current of biorthonormal version. Then $c$ is pumped charge
of all channel $k$ driven by the time-dependent Hamiltonian varying in a
period.

This conclusion is obtained for a model Hamiltonian with translational
symmetry. Next, we consider this issue based on the edge states for the
chain system.\textbf{\ }Based on the exact expression of the edge states, we
have known that only $\delta $\ can change the distribution of the
probability. Specifically, for a given $V>0$, if we vary $\delta $\ from $%
\delta _{0}>0$\ to $-\delta _{0}$\ adiabatically, the state $\overline{A}_{%
\text{L}}\left\vert \text{Vac}\right\rangle $\ will evolves to $\overline{A}%
_{\text{R}}\left\vert \text{Vac}\right\rangle $. Then particle will
transport adiabatically from the rightmost of the chain to the leftmost.
Remarkably, one can get the same result from an alternative way. We consider
a system with additional weak tunneling $\kappa $\ between two ends. The
corresponding Hamiltonian has the form%
\begin{equation}
H_{\kappa }=H_{\mathrm{CH}}+\kappa M,  \label{H_cappa}
\end{equation}%
which is schematically illustrated in Fig. \ref{fig2} and the structure of
spectrum of $H_{\kappa }$ is plotted in Fig. \ref{fig2}(a). We note that the
weak tunneling $\kappa $\ hybridize two edge $\overline{A}_{\text{L}%
}\left\vert \text{Vac}\right\rangle $\ and to $\overline{A}_{\text{R}%
}\left\vert \text{Vac}\right\rangle $\ states when $\delta >0$\ and $V$\
around $0$. For a given $\delta >0$, if we vary $V$\ from $V_{0}>0$\ to $%
-V_{0}$\ adiabatically, the state $\overline{A}_{\text{L}}\left\vert \text{%
Vac}\right\rangle $\ will evolves to $\overline{A}_{\text{R}}\left\vert
\text{Vac}\right\rangle $. Then a particle will transport adiabatically form
the rightmost of the chain to the leftmost. Fig. \ref{fig2}(b) and (d),
illustrate this point. During this process, particle should pass through the
connection of two ends rather than the bulk region of the chain. This
transport can be detect by watching the current across two ends.\textbf{\ }%
The accumulated charge of the current should be integer.\textbf{\ }

Inspired by these results\ we expect that the hidden topology behind the
non-Hermitian model can be unveiled by the pumping charge with the respect
to the mid-gap edge mode rather than all the energy levels. As illustrated
in Fig. \ref{fig3}(a), the pumping charge is the sum of all the pumping
charges obtained by an infinite rectangular loops. The pumping charge of a
rectangular loop is simply determined by the positions of two vertical
sides. Therefore, we can conclude that the pumping charge of an edge state
for an arbitrary loop equals to the corresponding Chern number of the loop.

To characterize the charge transfer quantitatively, we developed the concept
of the biorthonormal current in a non-Hermitian tight-binding model. In the
following, only single-particle case is concerned. We begin with the rate of
change of the biorthonormal probability $\rho _{l}$\ at an arbitrary site $l$
for given eigenstate $\left( |\varphi \left( t\right) \rangle ,\left\vert
\eta \left( t\right) \right\rangle \right) $ at instant $t$, which can be
expressed as%
\begin{eqnarray}
\frac{d\rho _{l}}{dt} &=&\frac{d\langle \eta \left( t\right) |l\rangle
\langle l|\varphi \left( t\right) \rangle }{dt}  \notag \\
&=&\frac{1}{i}\langle \eta \left( t\right) |\left[ \left\vert l\right\rangle
\left\langle l\right\vert ,H\right] |\varphi \left( t\right) \rangle
\end{eqnarray}%
where $\left\vert l\right\rangle =a_{l}^{\dag }\left\vert \text{Vac}%
\right\rangle $. For the concerned Hamiltonian we have%
\begin{equation}
\frac{d\rho _{l}}{dt}=j_{l}-j_{l-1},
\end{equation}%
where%
\begin{equation}
j_{l}=\frac{1}{i}\langle \eta \left( t\right) |(\kappa _{l,l+1}\left\vert
l\right\rangle \left\langle l+1\right\vert -\kappa _{l+1,l}|l+1\rangle
\langle l|)|\varphi \left( t\right) \rangle .
\end{equation}%
The quantity $j_{l}$\ refers to the biorthonormal current across sites $l$\
and $l+1$. The accumulated charge passing the site $l$ during the period $T$
is%
\begin{equation}
Q_{l}=\int_{0}^{T}\left( j_{l}-j_{l-1}\right) dt.
\end{equation}%
We consider the case by taking $V=V_{0}\left( 1-2\omega t\right) $ with $%
\omega \ll 1$, and
\begin{equation}
|\varphi \left( 0\right) \rangle =\overline{A}_{\text{L}}\left\vert \text{Vac%
}\right\rangle \text{, }\left\vert \eta \left( 0\right) \right\rangle =A_{%
\text{L}}^{\dag }\left\vert \text{Vac}\right\rangle .
\end{equation}%
According to the analysis above, if $t$ varies from $0$ to $T=\omega ^{-1}$,
$Q_{l}$\ should be $1$ or $-1$, which is consistent with the direction of
the loop. To examine how the scheme works in practice, we simulate the
quasi-adiabatic process by computing the time evolution numerically for
finite system. In principle, for a given initial eigenstate $\left\vert \psi
\left( 0\right) \right\rangle $, the time evolved state under a Hamiltonian $%
H_{\kappa }\left( t\right) $ is%
\begin{equation}
\left\vert \Phi \left( t\right) \right\rangle =\mathcal{T}\{\exp
(-i\int_{0}^{t}H_{\kappa }\left( t\right) \mathrm{d}t)\left\vert \psi \left(
0\right) \right\rangle \},
\end{equation}%
where $\mathcal{T}$ is the time-ordered operator. In low speed limit $\omega
\rightarrow 0$, we have%
\begin{equation}
f\left( t\right) =\left\vert \langle \overline{\psi }\left( t\right)
\left\vert \Phi \left( t\right) \right\rangle \right\vert \rightarrow 1,
\end{equation}%
where $\left\langle \overline{\psi }\left( t\right) \right\vert $\ is the
corresponding instantaneous eigenstate of $H_{\kappa }^{\dagger }\left(
t\right) $. The computation is performed by using a uniform mesh in the time
discretization for the time-dependent Hamiltonian $H_{\mathrm{CH}}\left(
t\right) $. In order to demonstrate a quasi-adiabatic process, we keep $%
f\left( t\right) >0.9985$\ during the whole process by taking sufficient
small $\omega $ (see the case (a1, b1) in Fig. \ref{fig4}). Fig. \ref{fig4}
plots the simulations of particle current and the corresponding total
probability for several typical cases, in order to see to what extent the
process can be regarded as a quasi-adiabatic one. It shows that the obtained
dynamical quantities are in close agreement with the expected Chern number.

\section{Summary}

\label{Summary}

In summary, we have analyzed a one-dimensional non-Hermitian RM model that
exhibits the similar topological features of a Hermitian one within the
time-reversal symmetry-unbroken regions, in which the Berry connection,
Berry curvature, Chern number, current, and pumped charge, etc., are defined
in the context of biorthonormal inner product. We also examined the
dynamical signature for topological invariant, which is pumped charge driven
by the parameters. The underlying mechanism of our finding is that if a
non-Hermitian system is pseudo-Hermitian and the Hermitian counterpart of
its is topological within the unbroken symmetric region, such a
non-Hermitian system inherits the same topological characterization of the
counterpart.

\acknowledgments We acknowledge the support of the CNSF (Grant No. 11374163).

\end{document}